\documentclass[twoside,twocolumn,9pt]{article}
\usepackage{extsizes}
\usepackage[super,sort&compress,comma]{natbib} 
\usepackage[version=3]{mhchem}
\usepackage[left=1.5cm, right=1.5cm, top=1.785cm, bottom=2.0cm]{geometry}
\usepackage{balance}
\usepackage{amsmath}
\usepackage{mathptmx}
\usepackage{sectsty}
\usepackage{graphicx} 
\usepackage{lastpage}
\usepackage[format=plain,justification=justified,singlelinecheck=false,font={stretch=1.125,small,sf},labelfont=bf,labelsep=space]{caption}
\usepackage{float}
\usepackage{fancyhdr}
\usepackage{fnpos}
\usepackage{multirow}
\usepackage{comment}
\usepackage[english]{babel}
\addto{\captionsenglish}{%
  
}
\usepackage{array}
\usepackage{droidsans}
\usepackage{charter}
\usepackage[T1]{fontenc}
\usepackage[usenames,dvipsnames]{xcolor}
\usepackage{setspace}
\usepackage[compact]{titlesec}
\usepackage{hyperref}

\newcommand{\OH}{OH$^+$}

\newcommand{\wn}{cm$^{-1}$}

\usepackage{epstopdf}

\definecolor{cream}{RGB}{222,217,201}

\begin{document}

\pagestyle{fancy}
\thispagestyle{plain}
\fancypagestyle{plain}{
\renewcommand{\headrulewidth}{0pt}
}

\makeFNbottom
\makeatletter
\renewcommand\LARGE{\@setfontsize\LARGE{15pt}{17}}
\renewcommand\Large{\@setfontsize\Large{12pt}{14}}
\renewcommand\large{\@setfontsize\large{10pt}{12}}
\renewcommand\footnotesize{\@setfontsize\footnotesize{7pt}{10}}
\renewcommand\scriptsize{\@setfontsize\scriptsize{7pt}{7}}
\makeatother

\renewcommand{\thefootnote}{\fnsymbol{footnote}}
\renewcommand\footnoterule{\vspace*{1pt}%
\color{cream}\hrule width 3.5in height 0.4pt \color{black} \vspace*{5pt}} 
\setcounter{secnumdepth}{5}

\makeatletter 
\renewcommand\@biblabel[1]{#1}            
\renewcommand\@makefntext[1]%
{\noindent\makebox[0pt][r]{\@thefnmark\,}#1}
\makeatother 
\renewcommand{\figurename}{\small{Fig.}~}
\sectionfont{\sffamily\Large}
\subsectionfont{\normalsize}
\subsubsectionfont{\bf}
\setstretch{1.125} 
\setlength{\skip\footins}{0.8cm}
\setlength{\footnotesep}{0.25cm}
\setlength{\jot}{10pt}
\titlespacing*{\section}{0pt}{4pt}{4pt}
\titlespacing*{\subsection}{0pt}{15pt}{1pt}

\fancyfoot{}
\fancyfoot[LO,RE]{\vspace{-7.1pt}\includegraphics[height=9pt]{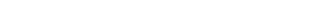}}
\fancyfoot[CO]{\vspace{-7.1pt}\hspace{13.2cm}\includegraphics{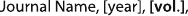}}
\fancyfoot[CE]{\vspace{-7.2pt}\hspace{-14.2cm}\includegraphics{head_foot/RF}}
\fancyfoot[RO]{\footnotesize{\sffamily{1--\pageref{LastPage} ~\textbar  \hspace{2pt}\thepage}}}
\fancyfoot[LE]{\footnotesize{\sffamily{\thepage~\textbar\hspace{3.45cm} 1--\pageref{LastPage}}}}
\fancyhead{}
\renewcommand{\headrulewidth}{0pt} 
\renewcommand{\footrulewidth}{0pt}
\setlength{\arrayrulewidth}{1pt}
\setlength{\columnsep}{6.5mm}
\setlength\bibsep{1pt}

\makeatletter 
\newlength{\figrulesep} 
\setlength{\figrulesep}{0.5\textfloatsep} 

\newcommand{\topfigrule}{\vspace*{-1pt}%
\noindent{\color{cream}\rule[-\figrulesep]{\columnwidth}{1.5pt}} }

\newcommand{\botfigrule}{\vspace*{-2pt}%
\noindent{\color{cream}\rule[\figrulesep]{\columnwidth}{1.5pt}} }

\newcommand{\dblfigrule}{\vspace*{-1pt}%
\noindent{\color{cream}\rule[-\figrulesep]{\textwidth}{1.5pt}} }

\makeatother

\twocolumn[
  \begin{@twocolumnfalse}
{\includegraphics[height=30pt]{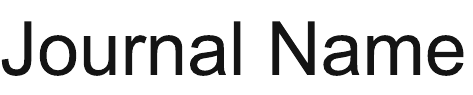}\hfill\raisebox{0pt}[0pt][0pt]{\includegraphics[height=55pt]{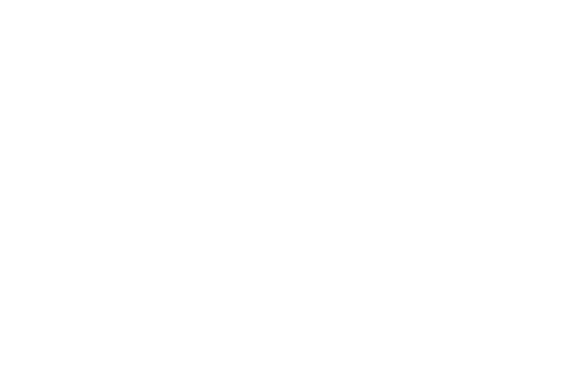}}\\[1ex]
\includegraphics[width=18.5cm]{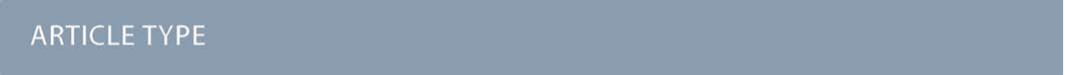}}\par
\vspace{1em}
\sffamily
\begin{tabular}{m{4.5cm} p{13.5cm} }

\includegraphics{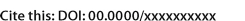} & \noindent\LARGE{
\textbf{
Hyperfine-resolved rovibrational and rotational spectroscopy of \OH  ($X~^3\Sigma^-$)
}} \\
 & \vspace{0.3cm} \\

 & \noindent\large{
Weslley G. D. P. Silva,\textit{$^{a}$} 
Lea Schneider,\textit{$^{a}$} 
Urs U. Graf,\textit{$^{a}$}  
Holger S.~P. Müller,\textit{$^{a}$} 
Pavol Jusko,\textit{$^{a,b}$}
Arshia M. Jacob,\textit{$^{a, c}$}  
Dominik Riechers,\textit{$^{a}$}
Stephan Schlemmer\textit{$^{a}$} and 
Oskar Asvany,\textit{$^{a}$}
 } \\

\includegraphics{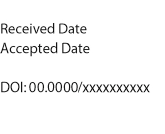} & \noindent\normalsize{
The \OH\ ($X~^3\Sigma^-$) radical cation has been investigated by  
combining a 4~K 22-pole ion trap apparatus with high-resolution IR and THz radiation sources.
Applying different types of action spectroscopic methods, the fundamental vibrational band in the 3~$\mu$m range and the spin manifold of the $N=1 \leftarrow 0$  rotational transition around 1~THz have been extended and refined. 
\textcolor{black}{Additionally,} the spin manifold of the $N=2 \leftarrow 1$ rotational transition, scattered around 2~THz, has been measured for the first time \textcolor{black}{with microwave accuracy}.
Although all \textcolor{black}{hyperfine components of the pure rotational transitions} are affected by considerable Zeeman splittings, a simulation of their contours allowed us to extract the field-free center frequencies with high accuracy. 
\textcolor{black}{A global fit combining rovibrational and pure rotational transitions from the literature with those newly obtained in this work was performed, leading to improvements in the spectroscopic constants of \OH, particularly those in the ground vibrational state.}
}

\end{tabular}

\end{@twocolumnfalse} \vspace{0.6cm}

  ]

\renewcommand*\rmdefault{bch}\normalfont\upshape
\rmfamily
\section*{}
\vspace{-1cm}


\footnotetext{\textit{$^{a}$~I. Physikalisches Institut, 
Universit\"at zu K\"oln, Z\"ulpicher Str.~77, 50937 K\"oln, Germany}\\
\textit{$^{b}$ Present address: CAS, Max-Planck-Institut für Extraterristrische Physik, 85741 Garching, Germany}}
\footnotetext{\textit{$^{c}$~Max-Planck-Institut für Radioastronomie, Auf dem Hügel 69, 53121 Bonn, Germany}}




\rmfamily 


\section{Introduction}
\label{intro}

Oxoniumylidene, \OH, is a simple radical cation with a $X~^3\Sigma^-$ electronic ground state.
The first observation of a rotational transition of this molecule was the $N = 1 \leftarrow 0$ multiplet which was studied in the laboratory by \citet{bee85} 
employing laser sideband spectroscopy, with rest frequencies calculated from the $A ^3\Pi - X ^3\Sigma^-$ electronic spectrum \citep{mer75}. 
Higher-$N$ transitions up to $N = 3$ were studied subsequently by laser magnetic resonance (LMR)\citep{gru86}, while the $N = 13 \leftarrow 12$ rotational transition was recorded by far-IR spectroscopy \citep{liu87a}.
The first high-resolution rovibrational spectrum in the IR was published by \citet{reh92} with vibrational excitation of hot \OH\ up to $v = 5 \leftarrow 4$. 
\citet{CDMS_2} presented a combined fit of field free data in the ground electronic state \citep{bee85,liu87a,reh92} which was the basis of rest frequencies used to search for \OH\ in space for a considerable time. 
Later, using advanced frequency-comb technology, \citet{mar16} determined \textcolor{black}{more} accurate ($\sim$MHz) rest frequencies of the \OH\ $v = 1 \leftarrow 0$ fundamental IR band, which were appended to the combined fit to improve the spectroscopic parameters.

In an astrochemical context, early chemical models predicted that \OH\ should be abundant in low-density interstellar environments, with its production driven 
by energetic processes such as interstellar shocks \citep{Glassgold1976, Barsuhn1977, deAlmeida1990}. 
Tentative observational support for this picture was reported in studies of shocked regions, including early claims of \OH\ in cometary bow shocks \citep{Balsiger1986}, although these detections were not definitive. 
Direct confirmation of widespread interstellar \OH\ remained challenging for several decades, largely because its ground-state rotational transition lies near 1~THz, 
a frequency range that is difficult to access from the ground owing to low atmospheric transmission, typically requiring exceptional weather conditions or space-/air-borne facilities.

Unambiguous detection of interstellar \OH\ was achieved only in 2010 through observations of the hyperfine components of its $N_J = 1_0 \leftarrow 0_1$ transition near 909~GHz \citep{wyr10} with the APEX telescope. 
This breakthrough demonstrated that \OH\ is indeed present in diffuse and translucent regions of the interstellar medium (ISM) and opened the door to systematic studies. 
Subsequent observations using space- and air-borne platforms, most notably the \textit{Herschel} Space Observatory and SOFIA (Stratospheric Observatory for Infrared Astronomy), enabled access to multiple low-lying rotational transitions 
and revealed \OH\ in both absorption and emission across a broad range of interstellar environments \citep{gerin10, Indriolo2015, Jacob2020, barlow13, vdT13, Jacob2022}. 
In parallel, \OH\ was also studied at ultraviolet wavelengths through its $A ^3\Pi - X ^3\Sigma^-$ electronic system \citep{mer75,Hodges2017}, 
providing complementary access to its spectroscopy and abundance in diffuse gas \citep{kre2010,Porras2014,Zhao2015,Bacalla2019}. 
These measurements established \OH\ as a widespread constituent of the ISM rather than a molecule confined to extreme or highly localized regions.
The close coupling between \OH\ chemistry and the ionization balance makes \OH, especially when considered alongside related species such as H$_2$O$^+$, a sensitive probe of the cosmic-ray ionization rate in diffuse atomic gas with molecular fractions $\leq 0.1$ 
(see Refs.~\citenum{Indriolo2015, neuf16, Jacob2020, kal23}).

Extragalactic detections of \OH\ have further demonstrated that its sensitivity to 
cosmic-ray ionization-driven chemical pathways is not unique to the Milky Way, 
but persists across a broad range of physical conditions and cosmic epochs, 
from nearby active and star-forming galaxies to the early universe \citep{vdW10, vdT2016, gon13, Muller2016, rie13, Indriolo2018}. 
At higher redshifts, the \OH\ $N = 1 \leftarrow 0$ transitions are shifted into 
frequency ranges accessible to sensitive (sub-)mm interferometers 
such as ALMA (Atacama Large Millimeter Array) and NOEMA  (Northern Extended Millimeter Array). 
Following its first detection in the highly redshifted 
 starburst galaxy, HFLS3 \citep{rie13} with a redshift of $z = 6.34$  \textcolor{black}{($z =(f_{\rm emit}-f_{\rm obs})/f_{\rm obs}$)}, 
at least one hyperfine component of \OH\ has now been detected in more than 
20 predominantly strongly lensed starburst galaxies at $z = 1.8$–6.3, 
enabling constraints on the ionization conditions in these intensely star-forming systems.

There have been fewer observations of higher-$N$ \OH\ transitions thus far, including toward a small number of Galactic planetary nebulae \citep{aleman14}, and toward the nearby galaxies NGC~4418, Arp~220 \citep{gon13}, and Markarian~231 \cite{ga18}. 
Valuable information on the excitation conditions can be obtained from observing these lines in absorption, emission, or a mix of both (so-called P-Cygni profile). 
In this context, precise laboratory measurements of pure rotational transitions of \OH\ at higher frequencies, such as the $N= 2-1$ transition at 2~THz, are needed.

Therefore, we employed a cryogenic ion trap apparatus in combination with the powerful leak-out spectroscopy method (LOS~\cite{scm22}),
explained in section~\ref{exptl}, to revisit the fundamental vibrational mode of \OH\ (section~\ref{IR-results}) in the present study. 
Subsequently, a double-resonance method has been applied to perform the first field-free laboratory measurement of the $N=2 \leftarrow 1$ rotational transition of \OH\ (section~\ref{rot-results}).
In addition, data of the $N=1 \leftarrow 0$ rotational transition are presented in section~\ref{rot-results_LIICG}.
With our novel measurements, sixteen hyperfine components could be extracted from the rotational measurements, and a global fit based on all available data is presented in section~\ref{fitting}.


\section{Experiment}
\label{exptl}

The experiments of this study were carried out using one of Cologne's 22-pole cryogenic ion trapping instruments, COLTRAP~\cite{asv14}. 
Every second (1~Hz), a pulse of about twenty thousand OH$^+$ ions was generated in a storage ion source via electron impact ionization of H$_2$O vapor ($E_{e^-} \approx 45$~eV). 
These ions were then extracted from the source, selected in a quadrupole mass spectrometer for \emph{m/z} = 17, and injected into the 4~K 22-pole ion trap~\cite{asv10}.
Upon entering the trap, the ions were stopped and thermalized to the cryogenic temperature 
by collisions with helium buffer gas ($n \approx 10^{13}$ cm$^{-3}$), which was added to the trap continuously.
During the trapping time of several hundred milliseconds,
the stored \OH\ ion cloud was irradiated with an IR and/or THz beam.
The outcome of this interaction has been probed via action spectroscopy by releasing the ion cloud into a second quadrupole mass spectrometer and counting the ions in a high-efficiency ion counter. 
A spectrum of \OH\ (rovibrational or rotational) can be recorded by counting the ions as a function of the radiation frequency (IR or THz). 
The different action spectroscopic methods and radiation sources employed in the present work are described in detail below.

\subsection{Leak Out Spectroscopy (LOS)}

The rovibrational spectrum of the fundamental stretching vibration of \OH\ was first measured using the leak out spectroscopy (LOS) method, described in detail by \citet{scm22}.
In the last four years, LOS has developed into a mature technique, with numerous recent applications to astrophysically relevant cations~\cite{asv23,bas23,scl23,sil23,gup23,sil24b,ste24,dom24,sil24,sal25,scl25,gup25,jim26}. 
The trapped ions are excited, in most applications of LOS  vibrationally by using an IR beam, and after inelastic collisions with a neutral atomic/molecular partner, 
in this case Ne, which is pulsed additionally into the trap in a 1:3 Ne:He mixture, 
a fraction of this excitation energy is converted into kinetic energy of both collision partners.
The additional kinetic energy gained by the ions allows them to escape from the trap and fly toward the detector, 
where they are counted as a function of the IR wavenumber to produce the rovibrational spectrum.

\subsection{Double-resonance rotational spectroscopy (DR-LOS)}
Once the rovibrational spectrum was investigated, the pure rotational transitions of \OH\ were 
measured using a double-resonance scheme based on LOS, or DR-LOS for short\cite{asv23}. 
This method has been applied to measure the $N=2 \leftarrow 1$ rotational line near 2~THz.
Here, the IR excitation is tuned to the resonant frequency of a rovibrational transition of the molecule, 
which creates a constant rovibrational LOS-signal. This signal is then altered by a resonant THz excitation, which can either 
increase (DR-LOS peak) or decrease (DR-LOS dip) the rovibrational LOS signal. 
If both transitions (IR and THz) start from the same lower quantum state, a dip is observed in the LOS spectrum, 
as the addressed quantum state is being de-populated by the THz excitation. 
In contrast, if the lower state of the IR transition corresponds to the upper state of the THz transition, 
the LOS signal is amplified (peak), as the THz excitation increases the population of the initial state of the rovibrational transition.
In both cases, a change in the number of escaping ions can be measured as a function of the THz frequency, 
thus generating a rotational spectrum. 
A more thorough review of double resonance schemes can also be found in Refs.~\citenum{asv21d,asv23},
and application examples for the DR-LOS method include  HCCCO$^+$ \cite{asv23},  c-C$_3$H$_2$D$^+$ \cite{gup23}, 
H$_2$CCCH$^+$ \cite{sil23},  HCNH$^+$ \cite{sil24}, C$_3$H$^+$ \cite{bad25}, 
NCCO$^+$\cite{bas25}, and recently HCN$^+$~\cite{sil26}.

\begin{figure*}[t]
\centering
  \includegraphics[width=1.0\textwidth]{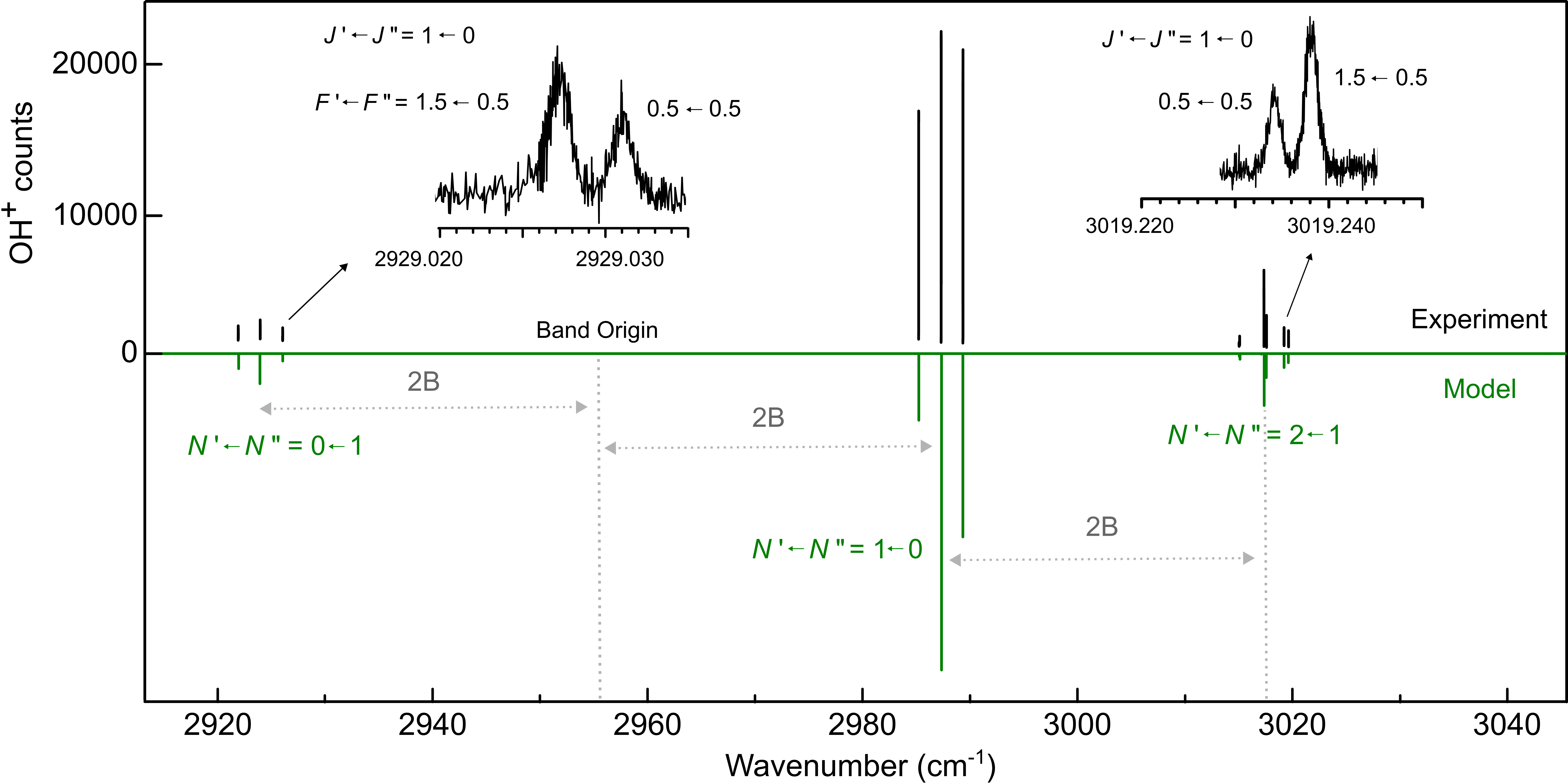}
  \caption{
The fundamental vibrational mode of \OH\ measured with leak-out spectroscopy (LOS).
The upper panel (in black) shows the measurement while the lower panel (in green) presents a simulation performed with PGOPHER~\cite{wes17} 
at a temperature of 20~K using the fitted spectroscopic parameters from Table~\ref{spec-parameter}. 
\textcolor{black}{The measurement displayed is "single-shot" with no averaging involved.}
The insets highlight the $N' \leftarrow N'' = 0 \leftarrow 1$, $J' \leftarrow J'' = 1 \leftarrow 0$ (left) 
and $N' \leftarrow N'' = 2 \leftarrow 1$, $J' \leftarrow J'' = 1 \leftarrow 0$ (right) rovibrational transitions with their resolvable 
$F' \leftarrow F'' = 0.5 \leftarrow 0.5$ and $F' \leftarrow F'' = 1.5 \leftarrow 0.5$ hyperfine components in detail.
\textcolor{black}{Due to the limited number of trapped \OH\ ions (about 20,000),
the strong lines in the $N' \leftarrow N'' = 1 \leftarrow 0$ branch saturate and therefore, their 
intensity pattern deviates from the simulation.}
}
  \label{IR_spec}
\end{figure*}


\subsection{Light Induced Inhibition of Complex Growth  (LIICG)}
The  rotational measurements \textcolor{black}{of the} $N=1 \leftarrow 0$ \textcolor{black}{transition} of \OH\  at 1~THz      
were carried out  using a method called light induced inhibition of complex growth (LIICG\cite{cha13,bru14}).
With LIICG, the rotational excitation of the molecular ion reduces the three-body rate  for attaching a He atom to the ion. 
Therefore, this method is performed by filling the trap with a larger number density of He gas ($\sim10^{15}$~cm$^{-3}$) for the He-attachment to proceed,
and by counting the number of product ions \OH--He (\emph{m/z} = 21) after the trapping period.
The resonant transition signal is then detected as a decrease (dip) in the \OH--He counts.
Before the invention of DR-LOS, LIICG as a generally applicable method 
has been extensively used for the rotational spectroscopy of astrophysically relevant 
cations~\cite{bru14,sto16,bru17,dom17,jus17,dom18,tho19,mar19,asv20b,asv21b}.

\subsection{Radiation Sources}

The narrow \textcolor{black}{linewidth} ($ < 10^{-4}$~\wn) tunable IR 
excitation was supplied by a continuous wave 
optical parametric oscillator (cw-OPO, Toptica, model TOPO) 
operating in the 2.5-3.5~$\mu$m spectral region. 
The power of the OPO admitted to the trap was on the order of a few tens of mW. 
The frequency of the IR radiation was measured continuously by a 
wavemeter/spectrum analyzer (Bristol Instruments, model 771A-MIR), 
which has an absolute accuracy of 0.2~ppm, i.e.,\ about 0.001~\wn.
As the 2~THz radiation source, operating in the range  1.83 -- 2.07~THz, the former local oscillator (LO)
of the Low Frequency Array (LFA) of the upGREAT receiver~\cite{ris16} onboard the SOFIA \cite{you12} air-borne observatory was used. 
In order to overlay the 2~THz radiation with the IR beam, a 1~mm hole in a plane mirror was used. 
The details of this setup have been explained elsewhere \cite{asv26}.
An amplifier-multiplier chain (Virginia Diode Inc., VDI) was applied as the 1~THz source. 
Both THz sources were driven by a synthesizer locked to a rubidium atomic clock.


\section{Spectroscopic properties of oxoniumylidene}
\label{spec-prop}

The \OH\ radical possesses two unpaired electrons ($S = 1$) and a $^3\Sigma^-$ ground electronic 
state which causes all rotational levels with $N > 0$ to split into three fine structure (FS) levels. 
The nuclear spin of H ($I = 1/2$) leads to hyperfine structure (HFS) which additionally splits each FS level into two. 
The rotational spacing is much larger than the FS splitting, which in turn is much larger than the HFS effects, so \OH\ is best described by Hund's case (b). 
The rotational angular momentum \textbf{N}, the electron spin angular momentum \textbf{S} and 
the nuclear angular momentum \textbf{I} are coupled sequentially as \textbf{J} = \textbf{N} + \textbf{S} and \textbf{F} = \textbf{J} + \textbf{I}. 

The selection rules are $\Delta N = \pm1$, $\Delta J = 0, \pm1$  and $\Delta F = 0, \pm1$. 
The most intense allowed transitions are those in which neither the electron spin nor the nuclear spin changes; 
thus $\Delta N = \Delta J = \Delta F$. 
Transitions with change in electron or nuclear spin are also allowed; 
their intensities can be substantial in comparison to the strongly allowed transitions for low quantum numbers, 
but are usually negligible for high quantum numbers. 
Moreover, such transitions usually carry more information on the FS or HFS parameters and are more susceptible to external magnetic fields, 
such as that of Earth, through the Zeeman effect.

The permanent electric dipole moment of \OH\ has not been determined experimentally to the best of our knowledge. 
A ground state value of $\mu_0=2.32$~D was estimated based on quantum chemical calculations~\citep{che07}.
Likewise, a $v = 1 - 0$ transition dipole moment of 0.185~D for the fundamental vibration has been reported previously~\citep{wer83}. 
It should be pointed out that the transition dipole moment alone is not always sufficient to model the IR intensities, 
as shown for the $v = 1 - 0$ band of CH$^+$ \citep{cha21}.


\begin{table*}[h]
  \caption{Observed  transitions of the fundamental vibrational band $v=1 \leftarrow 0$ of \OH\ (in \wn). 
  We calibrated our frequencies to the highly accurate lines of \citet{mar16} which are also given here.
}
  \label{IR_linelist}
  \begin{tabular*}{0.98\textwidth}{ccccccr@{}l}
    \hline
$N' \leftarrow N''$ &  $J' \leftarrow J''$ &   $F' \leftarrow F''$    &   this work     &                 & \citet{mar16}    & \multicolumn{2}{c}{obs$-$calc} \\
   \hline
0 $\leftarrow$ 1     & 1 $\leftarrow$ 1    &   1.5 $\leftarrow$ 1.5,  1.5 $\leftarrow$ 0.5 & 2921.8960(3) &     &                 &    0&.00026 \\ 
                     &                     &   0.5 $\leftarrow$ 0.5,  0.5 $\leftarrow$ 1.5 & 2921.9000(3) &     &                 &    0&.00034 \\
                     & 1 $\leftarrow$ 2    &                1.5 $\leftarrow$ 1.5           & 2923.9369(3) &     &                 & $-$0&.00021 \\ 
                     &                     &   1.5 $\leftarrow$ 2.5,  0.5 $\leftarrow$ 1.5 &              &     & 2923.940929(63) &    0&.000006\\
                     & 1 $\leftarrow$ 0    &                1.5 $\leftarrow$ 0.5           & 2926.0310(4) &     &                 &    0&.00043 \\
                     &                     &                0.5 $\leftarrow$ 0.5           & 2926.0348(4) &     &                 &    0&.00042 \\
1 $\leftarrow$ 0     & 0 $\leftarrow$ 1    &       0.5 $\leftarrow$ 0.5                    & 2985.2294(4) &     &                 & $-$0&.00044 \\ 
                     &                     &       0.5 $\leftarrow$ 1.5                    & 2985.2332(4) &     &                 & $-$0&.00044 \\
                     & 2 $\leftarrow$ 1    &   2.5 $\leftarrow$ 1.5, 1.5 $\leftarrow$ 0.5  &              &     & 2987.324274(96) &    0&.000008\\
                     &                     &            1.5 $\leftarrow$ 1.5               & 2987.3280(1) &     &                 & $-$0&.00009 \\
                     & 1 $\leftarrow$ 1    &   0.5 $\leftarrow$ 0.5, 1.5 $\leftarrow$ 0.5  & 2989.3492(1) &     &                 &    0&.00010 \\
                     &                     &   1.5 $\leftarrow$ 1.5, 0.5 $\leftarrow$ 1.5  &              &     & 2989.353127(63) &    0&.000116\\ 
2 $\leftarrow$ 1     & 1 $\leftarrow$ 1    &   0.5 $\leftarrow$ 0.5, 0.5 $\leftarrow$ 1.5  & 3015.1026(2) &     &                 & $-$0&.00009 \\
                     &                     &   1.5 $\leftarrow$ 1.5, 1.5 $\leftarrow$ 0.5  & 3015.1065(1) &     &                 & $-$0&.00010 \\
                     & 3 $\leftarrow$ 2    &   3.5 $\leftarrow$ 2.5, 2.5 $\leftarrow$ 1.5  &              &     & 3017.368784(61) &    0&.000024\\
                     &                     &         2.5 $\leftarrow$ 2.5                  & 3017.3726(1) &     &                 & $-$0&.00006 \\
                     & 2 $\leftarrow$ 1    &  2.5 $\leftarrow$ 1.5, 1.5 $\leftarrow$ 0.5, 1.5 $\leftarrow$ 1.5 &&& 3017.612933(52) &    0&.000066\\
                     & 1 $\leftarrow$ 0    &              0.5 $\leftarrow$ 0.5             & 3019.2375(1) &     &                 &    0&.00009 \\
                     &                     &              1.5 $\leftarrow$ 0.5             &              &     & 3019.241547(93) &    0&.000112\\
                     & 2 $\leftarrow$ 2    &   1.5 $\leftarrow$ 1.5, 2.5 $\leftarrow$ 1.5  & 3019.6542(1) &     &                 &    0&.00007 \\
                     &                     &   2.5 $\leftarrow$ 2.5, 1.5 $\leftarrow$ 2.5  & 3019.6580(1) &     &                 & $-$0&.00007 \\   
\hline
 \end{tabular*}
  {\scriptsize
\begin{flushleft}
\end{flushleft}
}
\end{table*}

\begin{table*}[h]
  \caption{Observed frequencies of the $N=1\leftarrow0$ and $N=2\leftarrow1$ rotational transitions of \OH\ (in MHz).
  Uncertainties are given in parentheses.}
  \label{linelist}
  \begin{tabular*}{0.88\textwidth}{cccr@{}lr@{}lr@{}l}
    \hline
$N' \leftarrow N''$   & $J' \leftarrow J''$ &   $F' \leftarrow F''$   &  \multicolumn{2}{c}{this work}     & 
\multicolumn{2}{c}{\citet{bee85}}  & \multicolumn{2}{c}{obs$-$calc} \\
   \hline
1 $\leftarrow$ 0  &  0 $\leftarrow$ 1    &   0.5 $\leftarrow$ 0.5  &   909045&.00(10)    & 909045&.2(10)  &   0&.044 \\
                  &                      &   0.5 $\leftarrow$ 1.5  &   909158&.88(10)    & 909158&.8(10)  &$-$0&.015 \\
                  &   2 $\leftarrow$ 1   &   2.5 $\leftarrow$ 1.5  &   971804&.43(10)    & 971803&.8(15)  &   0&.119 \\
                  &                      &   1.5 $\leftarrow$ 0.5  &   971805&.47(10)    & 971805&.3(15)  &   0&.015 \\
                  &                      &   1.5 $\leftarrow$ 1.5  &   971919&.46(10)    & 971919&.2(10)  &   0&.065 \\
                  &   1 $\leftarrow$ 1   &   0.5 $\leftarrow$ 0.5  &  1032998&.36(20)    & $-$6&.5(3)$^a$ &   0&.081 \\
                  &                      &   1.5 $\leftarrow$ 0.5  &  1033005&.11(10)    &1033004&.4(10)  &   0&.047 \\
                  &                      &   0.5 $\leftarrow$ 1.5  &  1033112&.25(20)    & $-$6&.8(4)$^a$ &   0&.031 \\
                  &                      &   1.5 $\leftarrow$ 1.5  &  1033119&.06(10)    &1033118&.6(10)  &   0&.058 \\
\hline
2 $\leftarrow$ 1  & 1 $\leftarrow$ 1     &   $^c$ \\
                  &  3 $\leftarrow$ 2    &    3.5 $\leftarrow$ 2.5  &  1959561&.9(1)$^b$  &  &   &    0&.138$^b$ \\
                  &                      &    2.5 $\leftarrow$ 1.5  &  1959561&.9(1)$^b$  &  &   & $-$0&.313$^b$ \\
                  &  2 $\leftarrow$ 1    &    1.5 $\leftarrow$ 1.5  &  1967532&.4(2)      &  &   & $-$0&.029 \\
                  &                      &    2.5 $\leftarrow$ 1.5  &  1967536&.1(1)      &  &   & $-$0&.094 \\
                  &                      &    1.5 $\leftarrow$ 0.5  &  1967539&.2(1)      &  &   & $-$0&.013 \\
                  &  1 $\leftarrow$ 0    &    0.5 $\leftarrow$ 0.5  &  2016070&.9(2)      &  &   & $-$0&.119 \\
                  &                      &    1.5 $\leftarrow$ 0.5  &  2016191&.6(2)      &  &   & $-$0&.147 \\
                  &  2 $\leftarrow$ 2    &    $^d$ \\
    \hline
  \end{tabular*}
  {\scriptsize
\begin{flushleft}
$^a$ \citet{bee85} reported frequency intervals between hyperfine transitions \\
$^b$ peaks merged; obs$-$calc of merged line is $-$0.048~MHz. \\
$^c$ Hyperfine lines weak and diluted by Zeeman effect. These components have not been targeted \\
$^d$ Hyperfine lines weak and diluted by Zeeman effect.  Only tentative detection
\end{flushleft}
}
\end{table*}

\begin{table}[h]
\begin{center}
\caption{Global fit Dunham-type spectroscopic parameters$^{a}$ of oxoniumylidene, \OH, from the present work in comparison to previous values.}
\label{spec-parameter}
\begin{tabular}[t]{lr@{}lr@{}l}
\hline \hline
Parameter                     & \multicolumn{2}{c}{This work} & \multicolumn{2}{c}{\citet{mar16}} \\
\hline
$Y_{10}$$^b$                  &            3119&.2892~(56)    &             3119&.2892~(56)  \\
$Y_{20}$$^b$                  &           $-$83&.1273~(57)    &            $-$83&.1273~(57)  \\
$Y_{30}$$^b$                  &               1&.01953~(241)  &                1&.01953~(241)\\
$Y_{40} \times 10^3$$^c$      &               2&.435~(453)    &                2&.435~(453)  \\
$Y_{50} \times 10^3$$^c$      &            $-$0&.241~(31)     &             $-$0&.241~(31)   \\
$Y_{01}$                      &          503486&.85~(22)      &           503486&.90~(26)    \\
$Y_{11}$                      &        $-$22435&.77~(62)      &         $-$22435&.77~(62)    \\
$Y_{21}$                      &             308&.47~(39)      &              308&.47~(39)    \\
$Y_{31}$$^c$                  &               1&.410~(51)     &                1&.410~(51)   \\
$Y_{02}$                      &           $-$58&.3435~(29)    &            $-$58&.3436~(59)  \\
$Y_{12}$                      &               1&.4529~(33)    &                1&.4523~(37)  \\
$Y_{22} \times 10^3$          &               7&.54~(135)     &                7&.51~(137)   \\
$Y_{03} \times 10^3$          &               4&.118~(20)     &                4&.115~(31)   \\
$Y_{13} \times 10^3$          &            $-$0&.1359~(155)   &             $-$0&.1326~(187) \\
$\lambda _{00}$               &           64379&.4~(25)       &            64379&.5~(26)     \\
$\lambda _{10}$               &          $-$253&.8~(66)       &           $-$254&.3~(66)     \\
$\lambda _{20}$               &           $-$24&.1~(33)       &            $-$24&.1~(33)     \\
$\lambda _{01}$               &            $-$0&.781~(24)     &             $-$0&.68~(11)    \\
$\gamma _{00}$                &         $-$4604&.96~(25)      &          $-$4605&.16~(42)    \\
$\gamma _{10}$                &             143&.41~(67)      &              143&.47~(67)    \\
$\gamma _{20}$                &            $-$1&.42~(35)      &             $-$1&.45~(35)    \\
$\gamma _{01}$                &               0&.806~(9)      &                0&.796~(17)   \\
$\gamma _{11}$                &            $-$0&.0385~(76)    &             $-$0&.0354~(88)  \\
$b_{F,00}(^1$H)               &           $-$74&.80~(5)       &            $-$75&.11~(49)    \\
$c_{00}(^1$H)                 &             125&.85~(20)      &              125&.95~(87)    \\
\textcolor{black}{$N$ $^d$}     & \textcolor{black}{365}&         & \textcolor{black}{328}&   \\
\textcolor{black}{rms error $^e$}                 &                   0&\textcolor{black}{.904}         & 0&\textcolor{black}{.908}    \\
\hline
\end{tabular}\\[2pt]
\end{center}
\footnotesize{
$^a$ Numbers in parentheses are one standard deviation in units of the least significant digits. 
All values are in units of MHz unless stated otherwise.\\
$^b$ In units of cm$^{-1}$.\\
$^c$ Sign of $Y_{31}$ was incorrect previously\cite{CDMS_2,mar16}.\\
$^d$ \textcolor{black}{Number of lines in the fit}.\\
$^e$ \textcolor{black}{Root mean square of the errors between the observed and calculated frequencies normalized to their uncertainties}.
}
\end{table}


\begin{table}[h]
\begin{center}
\caption{Ground state ($v=0$) spectroscopic parameters$^{a}$ of oxoniumylidene, \OH, from the present work in comparison to values from a previous two-state fit \cite{mar16}.}
\label{spec-parameter2}
\begin{tabular}[t]{lr@{}lr@{}l}
\hline \hline
Parameter       & \multicolumn{2}{c}{This work} & \multicolumn{2}{c}{\citet{mar16}} \\
\hline
$B_0$                  &  492346&.300~(28)      &        492346&.278~(146) \\
$D_0$                  &      57&.6232~(45)     &            57&.6166~(52)   \\
$H_0$                  &       0&.004075~(30)   &             0&.004049~(26) \\
$\lambda_0$            &   64246&.45~(7)        &         64246&.00~(55)    \\
$\lambda_{D0}$         &    $-$0&.77~(3)        &          $-$0&.54~(12)     \\
$\gamma_0$             & $-$4533&.60~(4)        &       $-$4533&.85~(34)    \\
$\gamma_{D0}$          &       0&.7889~(89)     &             0&.7847~(153)  \\
$b_F$($^1$H)           &   $-$74&.79~(5)        &         $-$75&.14~(50)    \\
$c$($^1$H)             &     125&.77~(20)       &           126&.01~(87)     \\
\hline
\end{tabular}\\[2pt]
\end{center}
\footnotesize{
$^a$ Numbers in parentheses are one standard deviation in units of the least significant digits. 
All values are in units of MHz.\\
}
\end{table}

\section{IR measurements}
\label{IR-results}

A prerequisite for performing double resonance rotational spectroscopy is the knowledge of the rovibrational transitions of the molecule under consideration. 
Therefore, we revisited the fundamental vibrational transitions of \OH\ which have been first pioneered by \citet{reh92} and later refined by frequency-comb-accuracy measurements by \citet{mar16}.
We used leak-out spectroscopy (LOS~\cite{scm22}) as \textcolor{black}{the} action spectroscopy method in the present study.
Our recorded LOS spectrum is shown in Fig.~\ref{IR_spec}. 
Only one $P$-branch ($N=0\leftarrow1$) and two $R$-branch manifolds ($N=1\leftarrow0$, $N=2\leftarrow1$) could be recorded because of the cryogenic cooling. 
Also, the low temperature resulted in very narrow Gaussian lines with about 45~MHz FWHM (full width at half maximum). 
We consider this FWHM to be caused solely by Doppler broadening, corresponding to a kinetic temperature of the \OH\ ions in the trap of about 12~K, a temperature very typical for our ion trapping experiments~\cite{koh18,dom18,asv23,scl23}.
Other broadening effects appear to be negligible.
The observed narrow lines made it possible to resolve some hyperfine components for the first time. 
An example is shown in the zoom-in panel given in Fig.~\ref{IR_spec} and the newly 
obtained components are tabulated in Table~\ref{IR_linelist}. 
%


\section{Rotational measurements: \texorpdfstring{$N=2 \leftarrow 1$}{N = 2 - 1}}
\label{rot-results}

\begin{figure*}[h]
\centering
  \includegraphics[width=1.0\textwidth]{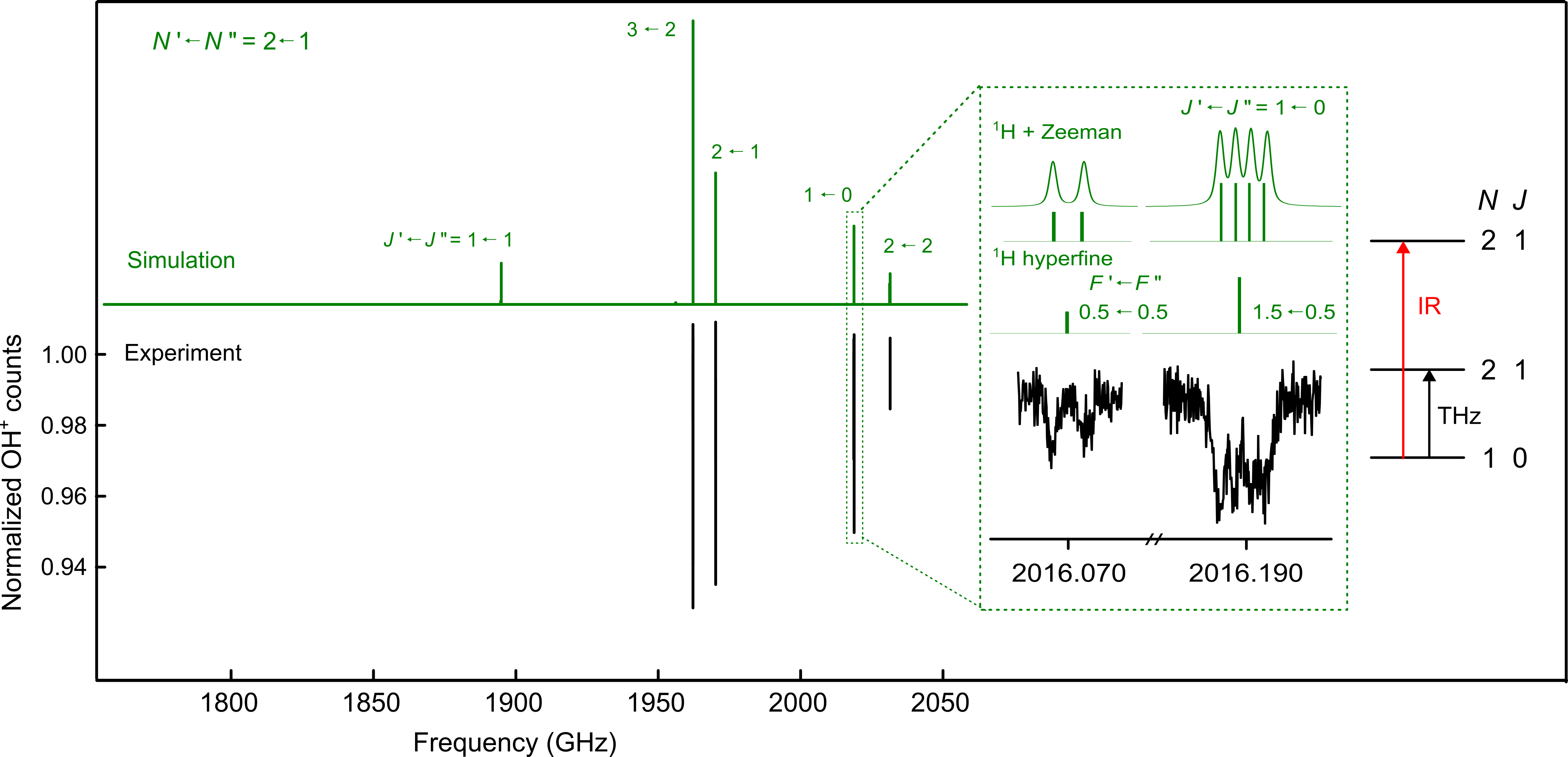}
  \caption{
Rotational spectroscopy of the $N= 2 \leftarrow 1$ spin manifold of \OH\ obtained using the DR-LOS "dip" method.
The upper panel shows the PGOPHER~\cite{wes17} simulation (in green), while the lower panel presents the measured spectrum (black). 
To measure the  $J= 1 \leftarrow 0$ rotational transition shown in the inset, for example, 
the IR beam was kept fixed on resonance with the rovibrational $(\nu,N,J)= (1,2,1) \leftarrow (0,1,0)$ transition at 3019.2375~\wn, 
while the THz excitation was used to de-populate the $(\nu,N,J)= (0,1,0)$ common level. 
This results in a decrease of rovibrational LOS signal, i.e.\ in the \OH\ counts, and the rotational transition is observed as a dip.
The inset also provides a detailed illustration of the hyperfine components, along with the effects of Zeeman splitting and Doppler broadening, 
to which the lines of \OH\ are subjected to owing to its open shell nature.
}
  \label{rot_spec}
\end{figure*}

The spin manifold of the $N=2 \leftarrow 1$ rotational transition of \OH\ was predicted to fall exactly in the tuning range of one of our THz sources (1.83-2.07~THz~\cite{ris16}).
The lower state quantum levels with $N=1$ lie at about 30-34.5~\wn, so it was expected that measuring these transitions \textcolor{black}{would} be challenging due to the low population of the levels at 4~K. 
To measure these lines with the IR-THz DR-LOS scheme~\cite{asv23}, IR transitions connecting to these rotational quantum levels were required,
e.g.\ $N=2 \leftarrow 1$ or $N=3 \leftarrow 2$. 
As the latter IR transitions were hardly detectable due to the negligible population of the $N=2$ ($\approx$100~\wn) states at low temperature,
we chose the $N=2 \leftarrow 1$ IR manifold 
\textcolor{black}{shown in Fig.~\ref{IR_spec}} for the double resonance method (or $N=0 \leftarrow 1$).
The signal is seen as a drop in the ion counts due to the DR configuration. 
The measured rotational spectrum is shown in Fig.~\ref{rot_spec}, and the inset shows the applied double resonance scheme.

The spectrum in Fig.~\ref{rot_spec} is a composite of several individual measurements, in which the IR frequency was kept fixed on the respective resonance, and the submm-wave frequency was scanned back-and-forth \textcolor{black}{multiple times} in a given frequency window in fixed steps \textcolor{black}{of 50~kHz}. 
This large step size was necessary because of the extended Zeeman features (see below). 
The signal in Fig.~\ref{rot_spec} was normalized following a frequency-switching procedure, where the \OH\ ion counts monitored in the frequency window of interest are divided by the \OH\ counts at an off-resonance frequency position. 
Thus, the baseline in the spectrum of Fig.~\ref{rot_spec} is close to unity.
\textcolor{black}{After normalization, the up-and-down scans were averaged into a  
single spectrum.}


\section{Rotational measurements: \texorpdfstring{$N=1 \leftarrow 0$}{N = 1 - 0}}
\label{rot-results_LIICG}


The spin manifold of the $N=1\leftarrow 0$ rotational transition around 1~THz was targeted with another multiplier chain radiation source (Virginia Diode Inc., VDI).
\citet{bee85} pioneered the measurement of these lines 40 years ago, and by revisiting these \textcolor{black}{transitions using our experimental setup at low temperature and low pressure}, it was possible to refine the line uncertainties by a factor of about ten.
We applied LIICG~\cite{bru14} as the action spectroscopic method, which is a (sub)mm-wave-only method without the need for a laser.
The data of these measurements were recorded employing the COLTRAP machine between December 2016 and January 2017, and are published here for the first time. 
The measurement procedure and frequency-switching normalization is very similar to that described in the preceding section, and is well documented in the literature~\cite{bru14,sto16,bru17,dom17,jus17,dom18,tho19,mar19,asv20b,asv21b}.


\section{Extraction of hyperfine line list}
\label{HFS-info}

The \OH\ ($X~^3\Sigma^-$) radical is susceptible to the Zeeman effect in magnetic fields due to its open shell nature.
This was already exploited by the LMR measurements of \OH\ by \citet{gru86}.
We observe a splitting of lines in our DR-LOS and LIICG measurements, which could be well reproduced 
by including the Zeeman effect in our PGOPHER~\cite{wes17} simulations, as shown in the inset in Fig.~\ref{rot_spec}. 
The magnetic field best reproducing the DR-LOS observations is on the order of 160~$\mu$T 
and thus about three times larger than the magnitude of the Earth's magnetic field here in Cologne. 
This value agrees well with recent measurements of other open shell species in our laboratory,
e.g. CCH$^+$ ($X~^3\Pi$)\cite{ste26} or HCN$^+$ ($X~^2\Pi$)\cite{sil26}, indicating that the observed Zeeman splitting may arise from our experimental setup, such as from magnetic parts of the vacuum chamber or the magnetically levitated turbo-pumps.
Interestingly, the magnetic field best reproducing the LIICG measurements from 2016/2017
is about 100~$\mu$T, which would favor the turbo-pump theory, as a change of pump type indeed happened in-between.
 
We simulated the lines with PGOPHER including the Zeeman effect and shifted the simulation in frequency until a good match with the experimental trace had been achieved to extract the positions of the field-free hyperfine components from our measured data.
We then obtained the line centers by setting the B-field in the simulation to zero.
The hyperfine components obtained this way are listed in Table~\ref{linelist}.
We estimate these field-free line positions to have an uncertainty on the order of 100--200~kHz.


\section{Global fit}
\label{fitting}

The \OH\ frequency data obtained in the course of the present investigation were \textcolor{black}{combined with the most complete dataset for \OH, which was latest reported in the work of \citet{mar16}}.
\textcolor{black}{This resulting linelist} was subjected to a fit to determine the spectroscopic parameters employing Pickett's SPFIT program \citep{pic91}. The starting values of the parameters were those from the \textcolor{black}{fit} of \citet{mar16}.
Briefly, the rovibrational energy levels of a diatomic molecule can be represented by the Dunham expression
\begin{equation}
\label{Dunham}
E(v, J)/h = \sum_{i,j} Y_{ij}(v + 1/2)^i J^j (J + 1)^j,
\end{equation}
where $Y_{ij}$ are the Dunham parameters. 
The electron spin-electron spin coupling parameters $\lambda _{ij}$, the electron spin-rotation parameter $\gamma _{ij}$ 
and the isotropic and anisotropic electron spin-nuclear spin coupling parameters $b_{F,ij}$ and $c_{ij}$, respectively, can be expanded in an equivalent way. 
It is worthwhile to mention that in the case of the last two parameters, only fundamental parameters 
with $i = j = 0$ were applied before\cite{mar16} and in the present work; vibrational corrections to the HFS parameters could not be determined with sufficient certainty. 
The resulting spectroscopic parameters are summarized in Table~\ref{spec-parameter}, \textcolor{black}{where the values from the fit of \citet{mar16} are also given for comparison. Overall, the rms
error of the present fit (0.904) is slightly smaller than the previously reported value (0.908).}


The line-, parameter- and fit-files of \OH\ along with auxiliary files will be made available in the 
data section\footnote{https://cdms.astro.uni-koeln.de/classic/predictions/daten/OH+/} 
of the Cologne Database for Molecular Spectroscopy , CDMS.\cite{CDMS_2,CDMS_3} 
In addition, the \OH\ $v = 0$ catalogue file in the entry 
section\footnote{https://cdms.astro.uni-koeln.de/classic/entries} of the CDMS will 
be updated and a new entry will be created for the $v = 1 - 0$ IR band. 
For completeness, the spectroscopic parameters for the ground state ($v=0$), obtained by a global fit 
including all field-free measurements involving that state~\cite{bee85,liu87a,reh92,mar16},
are given in Table~\ref{spec-parameter2}.



\section{Discussion}
\label{discussion}


The cryogenic and mass-selective ion-trapping approach~\cite{ger92,mcg20,asv21d} of this work made it possible to refine
the $N= 1 \leftarrow 0$ transitions of \OH\  at 1~THz, and to measure the  $N= 2 \leftarrow 1$ manifold at 2~THz for the first time, utilizing the local oscillator of the upGREAT instrument~\cite{ris16}.
The investigation was enabled by applying different action spectroscopic methods, LIIGC~\cite{bru14,bru17} and DR-LOS~\cite{asv23}.
In the latter approach, the rovibrational excitation following the rotational excitation leads to 
the ions being kicked out of the ion trap and being detected. 
At higher rotational frequencies, such as the 2~THz regime applied in this work, the rotational photon alone carries enough energy to eject the ions from the trap.
This has been demonstrated only very recently and is termed rot-LOS (rotational leak-out spectroscopy~\cite{asv26}). 
For the strongest transition \textcolor{black}{$N = 2 \leftarrow 1$}, $J = 3 \leftarrow 2$ of \OH\ at 1959.56~GHz, we tested rot-LOS and indeed obtained a strong signal. 
Such a rotational-only approach might be useful in the future when no other radiation source is available 
to perform a double-resonance, or \textcolor{black}{when} the rovibrational spectrum is not or only insufficiently known.


\textcolor{black}{Owing to the new data obtained in this work, the uncertainties in the values of several spectroscopic parameters of} \OH\, were reduced by varying degrees, \textcolor{black}{as summarized in Tables~\ref{spec-parameter} and \ref{spec-parameter2}}.
\textcolor{black}{In particular, the uncertainty in the ground-state rotational parameter $B_0$ decreased from 146 kHz in the previous fit of \citeauthor{mar16} to 28 kHz in the present work, while the $B_0$ values themselves remain in excellent agreement, differing by only 22 kHz (Table~\ref{spec-parameter2}).}
\textcolor{black}{Although} higher-order and purely vibrational parameters were largely unaffected, the uncertainties 
of $Y_{02}$, $Y_{03}$, $\lambda_{01}$ and $\gamma_{01}$ have been reduced by factors of $\sim$1.5 to $\sim$4 \textcolor{black}{(Table~\ref{spec-parameter})}. 
Notable are also the improvements in the HFS parameters $b_F$ and $c$ by factors of about 10 and 4, respectively. 

\textcolor{black}{Since \OH\ is isoelectronic with the NH radical, it is interesting to compare their HFS constants, particularly those of the $^1$H nucleus.}
In fact, $b_F$\textcolor{black}{(H)} and $c$\textcolor{black}{(H)} of NH were determined as $-66.131 \pm 0.015$~MHz and $90.291 \pm 0.084$~MHz, respectively,\cite{F-M04} slightly smaller in magnitude than \textcolor{black}{the corresponding values for \OH, which are $-74.80 \pm 0.05$~MHz and $125.85 \pm 0.20$~MHz (Table~\ref{spec-parameter})}.
These values may \textcolor{black}{also} be compared with those of their heavier siblings, PH ($-46.543 \pm 0.004$ MHz and $19.39 \pm 0.06$~MHz, respectively \cite{kli1998}) 
and SH$^+$ ($-56.84 \pm 0.03$~MHz and $33.48 \pm 0.13$~MHz, respectively\cite{Muller2017}). 
\textcolor{black}{The HFS parameters of PH and SH$^+$ differ not only from those of the lighter \OH\ and NH counterparts, but also among each other.}





With the laboratory measurements of the $N=2 \leftarrow 1$ transition now secured, the question turns to its astrophysical impact. 
As discussed in the introduction, detections of the $N=2 \leftarrow 1$ transitions were achieved toward three Galactic sources and three galaxies in the local universe, 
but due to the end of the {\em Herschel} mission, this observing opportunity no longer exists. 
However, these lines become redshifted into the observable window in the early universe. 
Indeed, \citet{rie13} show an observed-frame 1~mm spectrum of the $z=6.34$ starburst HFLS3, obtained with the Z-Spec instrument mounted on the Caltech Submillimeter Observatory (CSO). 
Based on similar line identifications in Arp\,220 at the time (i.e., before the {\em Herschel}/PACS spectrum was publicly available; \citealt{gon13}), 
two weak absorption features are identified as excited CH and NH transitions, but both are blended with OH$^+$ $2 \leftarrow 1$ lines at the spectral resolution of Z-Spec. 
Indeed, recent high-resolution spectroscopy with NOEMA reveals strong contributions of OH$^+$ $N=2 \leftarrow 1$ absorption to these blended features (D.~A. Riechers, priv. comm.). 
Analogous to the $N=1 \leftarrow 0$ detection in this source, the confirmation of $N=2 \leftarrow 1$ lines opens up the study of excited OH$^+$ in the early universe, 
as is key for the study of the strongest far-IR emitting regions in the circumnuclear regions of the most intense, massive starbursts known.


In conclusion, the investigation of \OH\  ($X~^3\Sigma^-$) in this work using LOS \textcolor{black}{reinforces} that this \textcolor{black}{novel} method is perfectly suited 
for investigating open shell cations, as \textcolor{black}{recently also} shown for CCH$^+$ ($X~^3\Pi$) \cite{ste26} and for the isomers 
HNC$^+$ ($X~^2\Sigma$) and HCN$^+$ ($X~^2\Pi$) \cite{jim26,scm26,sil26}. 
\textcolor{black}{While these} investigations have \textcolor{black}{primarily focused} on 
rovibrational and \textcolor{black}{rotational spectroscopy}, vibronic~\cite{mar25a} and rovibronic~\cite{mar26} LOS have also \textcolor{black}{recently} been demonstrated, \textcolor{black}{highlighting the broad applicability of LOS for investigating molecular cations across a wide range of wavelengths.}
\textcolor{black}{In this context, revisiting the $A ^3\Pi - X ^3\Sigma^-$ electronic spectrum of \OH\ \cite{mer75,Hodges2017} using LOS is a promising future experiment.}
\textcolor{black}{With the improved determination of the spectroscopic parameters achieved here, searches for rotational transitions of \OH\ at higher frequencies should also be facilitated.}

\section*{Conflicts of interest}
There are no conflicts to declare.

\section*{Data availability}
The original data of the IR measurement and the two rotational measurements (at 1 and 2~THz) are made available in the 
supplementary material as ASCII files together with explanations.
\textcolor{black}{In addition, the fit files used 
(PGOPHER as well as SPFIT/SPCAT) are also provided. }

\section*{Acknowledgements}
This work has been supported by an ERC Advanced Grant (MissIons: 101020583), and by the 
Deutsche Forschungsgemeinschaft (DFG) via the 
Collaborative Research Center SFB~1601 (project ID: 500700252, 
sub-projects B7, B8, C1, C2, C4 and Inf) and via SCHL 341/15-1 
(``Cologne Center for Terahertz Spectroscopy''). 
W.G.D.P.S.  and P.J. thank the Alexander von Humboldt Foundation for 
support through  postdoctoral fellowships. 
The Toptica cw-OPO has been financed by HBFG (INST 216/1184-1).

\newpage


\providecommand*{\mcitethebibliography}{\thebibliography}
\csname @ifundefined\endcsname{endmcitethebibliography}
{\let\endmcitethebibliography\endthebibliography}{}

\end{document}